\documentclass[%
 reprint,
 amsmath,amssymb,
 aps,
]{revtex4-2}

\usepackage{graphicx}
\usepackage{dcolumn}
\usepackage{bm}

\usepackage{pdfpages}  
\usepackage{pgffor}

\makeatletter
\AtBeginDocument{\let\LS@rot\@undefined}
\makeatother

\newcommand{\revision}[1]{\textcolor{black}{{#1}}}

\begin{document}

\preprint{arxiv}

\title{Giant segregation transition as origin of liquid metal embrittlement \\ in the Fe-Zn system}

\author{Reza Darvishi Kamachali$^{1}$} 
\email{Corresponding author: reza.kamachali@bam.de (Reza Darvishi Kamachali)}
\author{Theophilus Wallis$^1$} 
\author{Yuki Ikeda$^1$}
\author{Ujjal Saikia$^2$}
\author{Ali Ahmadian$^2$}
\author{Christian H. Liebscher$^2$}
\author{Tilmann Hickel$^{1,2}$}
\author{Robert Maa\ss$^{1,3}$}

\affiliation{$^1$\textit{Federal Institute for Materials Research and Testing (BAM), Unter den Eichen 87, 12205 Berlin, Germany}}
\affiliation{$^2$Max-Planck-Institut f\"ur Eisenforschung GmbH, Max-Planck-Stra{\ss}e 1, 40237 D\"usseldorf, Germany}
\affiliation{$^3$University of Illinois at Urbana-Champaign, Urbana, 61801 Illinois, USA}


\begin{abstract}
A giant Zn segregation transition is revealed using CALPHAD-integrated density-based modelling of segregation into Fe grain boundaries (GBs). The results show that above a threshold of only a few atomic percent Zn in the alloy, a substantial amount of up to 60 at.\% Zn can segregate to the GB. We found that the amount of segregation \revision{abruptly} increases with decreasing temperature, while the Zn content in the alloy required for triggering the segregation transition decreases. Direct evidence of the Zn segregation transition is obtained using high-resolution scanning transmission electron microscopy. Base on the model, we trace the origin of the segregation transition back to the low cohesive energy of Zn and a miscibility gap in Fe-Zn GB, arising from the magnetic ordering effect, which is confirmed by ab-initio calculations. We also show that the massive Zn segregation resulting from the segregation transition greatly assists with liquid wetting and reduces the work of separation along the GB. \revision{The current predictions suggest that control over Zn segregation, by both alloy design and optimizing the galvanization and welding processes, may offer preventive strategies against liquid metal embrittlement.}
\end{abstract}

\maketitle

Liquid Metal Embrittlement (LME) is a major safety concern in structural engineering.
For Zn-coated steels, LME leads to severe limitations in manufacturing and related designs in the automotive industry \cite{sierlinger2017cracking,bhattacharya2018liquid}. 
Extensive research has been devoted to understanding the underlying mechanisms of LME (see for instance \cite{joseph1999liquid,nilsson2018overview,razmpoosh2021pathway} and references therein).
Embedded in several analyses, wetting of grain boundaries (GBs) has been proposed as a leading mechanism in promoting crack initiation and propagation in Zn welding regions \cite{wolski2001importance,bhogireddy2016liquid}.
The extent and significance of wetting are argued to depend on the solubility of the embrittling metal in the base alloy \cite{shunk1974specificity}. 
Another mechanism repeatedly discussed for LME is stress-assisted diffusion, promoting the formation of a Zn-penetration-zone and thus cracking \cite{ling2020towards,digiovanni2021liquid}.
Recently, it has also been proposed that LME can be driven by the formation of brittle Zn-rich $\Gamma$ precipitates that further create stress heterogeneities in the uncracked GBs \cite{ikeda2022early}. 

As a common point of departure in various mechanisms, the interaction between Zn and Fe atoms at GBs is of fundamental importance: 
In particular, one can ask if, prior to wetting or GB precipitation, the segregation of Zn could play a critical role on a much smaller scale? 
Indeed, careful investigations revealed that Zn-rich precipitates form along the GB far ahead of the crack tip \cite{ikeda2022early,ikeda2023segregation}. 
Furthermore, atom probe tomography has revealed significant elemental segregation along GBs \cite{razmpoosh2021atomic}. 
These observations brace Zn segregation as a \textit{precursor} for LME, weakening the GB and accelerating interfacial wetting and precipitation prior to crack initiation. 
Even for the mechanism of stress-assisted diffusion, the enlargement of the so-called Zn-penetration-zone should be studied in conjunction with the GBs' segregation response. 
For these reasons and because of its potential for preventive design against LME, it is highly desirable to explore and understand Zn segregation and related GB phase behavior. 

In the bulk Fe-Zn phase diagram, the BCC solid solution shows a miscibility gap over a significant range of temperatures and compositions, see Fig. S1 in the Supplementary Material (SM).
Although often neglected in the usual Fe-Zn phase diagram (due to the assumption of an equilibrium formation of the Zn-rich $\Gamma$ phase), this miscibility gap has important implications for the Zn segregation behavior:
In fact, the presence of such a miscibility gap can lead to a segregation transition phenomenon as demonstrated in other alloy systems \cite{darvishikamachali2020model,darvishikamachali2020segregation,wang2021density}.
Furthermore, due to the small cohesive energy of Zn, the driving force for any Zn segregation is expected to be high, i.e., the energy of an Fe GB can be dramatically reduced with Zn segregation. 
To capture these two major features, density-based Gibbs free energy formalism for describing GBs is considered \cite{darvishikamachali2020model}: 
\begin{align}
G(T, \rho, X_{Zn})
&=  X_{Fe} G_{Fe} (T, \rho) + X_{Zn} G_{Zn} (T, \rho) \nonumber \\
&+ {\rho^2} \Delta H^B \left( T, X_{Zn} \right) - T \Delta S^B \left( T, X_{Zn} \right)
\label{eq_G_alloy}
\end{align}
with $G_i(T, \rho) = {\rho^2} E_i^B+{{\rho}} \left( G_i^B(T) -E_i^B \right)$.
Here, the dimensionless atomic density $\rho$ defines the parent bulk phase when $\rho = 1$ or a GB when $\rho < 1$. 

\newpage
\onecolumngrid 
\setlength{\belowcaptionskip}{-15pt}
\begin{figure*}[t]
    \centering
   \includegraphics[width=1.0\textwidth]{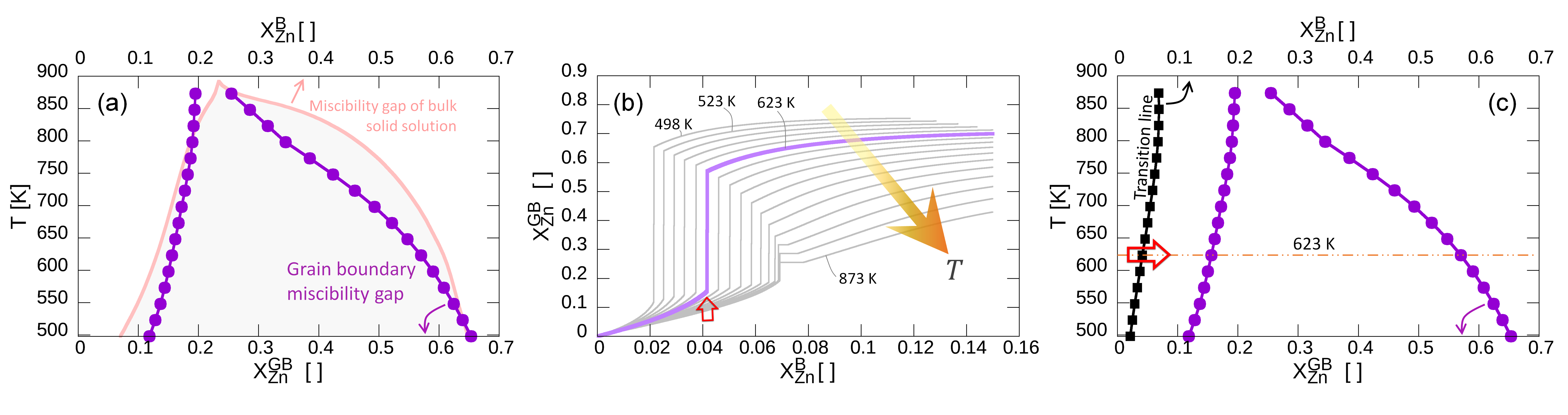}
    \caption[width=1.0\textwidth]{
    \textbf{Grain boundary (GB) phase diagram.}
    (a) The GB miscibility gap (circles) computed utilizing the CALPHAD-integrated density-based method. 
    Thermo-Calc TCFE11 database was used. 
    The miscibility gap of the parent $\alpha$ (BCC) bulk phase is shown for comparison in the background. 
    The superscripts B and GB refer to bulk and grain boundary compositions, respectively.
    (b) GB segregation isotherms computed for various temperatures using CALPHAD-integrated density-based method.
    The abrupt jump in each isotherm indicates the segregation transition. 
    (c) The complete density-based GB phase diagram is obtained, combining the results in panels (a) and (b).
    At each given temperature, the \textit{transition line} (squares) marks the critical alloy (bulk) composition at which the segregation transition occurs.
    Massive Zn segregation transition is predicted for alloy compositions and temperatures changing from the left to the right side of the transition line.
    The segregation transition at $T = 623$ K is marked, as an example, to link the results in panels (b) and (c).\\
    }
    \label{fig_GBPhDia}
\end{figure*}
\twocolumngrid 
 All other terms are bulk thermodynamic properties available from CALPHAD databases: 
The Gibbs free energy $G_i^B$ and potential energy $E_i^B$ of the pure $i$ substance, the mixing enthalpy $\Delta H^B$ and the mixing entropy $\Delta S^B$ ---the superscript $B$ denotes bulk properties throughout this paper.
 Compared to its original formulation \cite{darvishikamachali2020model}, in this study, we omit the gradient energy contributions in the free energy and, instead, focus on the temperature and composition dependence of the Zn segregation phenomenon. 
 
The density-based approach has been successfully applied for microstructure and alloy design in a broad range of materials \cite{li2020grain,wang2021incorporating,zhou2021spinodal,wallis2023grain,wang2023calphad}. 
The key strength of this method is its natural capacity to be integrated with CALPHAD framework and atomistic simulations, providing in-depth multi-scale insights into the phase behavior of GBs. 
We study a prototypical GB with $\rho = 0.92$ motivated by atomistic simulations of BBC-Fe GBs. 
For the calculations, we use the Thermo-Calc TCFE11 database.
The computational details and input parameters are given in Appendix A and B in the SM.

Figure \ref{fig_GBPhDia}(a) shows the computed GB miscibility gap (solid lines with circles). 
Compared to the parent bulk phase (red curve in the background), the GB miscibility gap slightly shrinks and shifts, however, it still covers a broad range of temperatures and compositions. 
Applying the equal chemical potential condition \cite{darvishikamachali2020model}, one can study the coexistence of the GB and bulk.
This gives the segregation isotherms shown in Fig. \ref{fig_GBPhDia}(b). 
The isotherms reveal a segregation transition, i.e., an abrupt increase (jump) in the GB Zn segregation with increasing Zn content in the bulk solid solution.
This corresponds to the miscibility gap of the GB.
We found that the amount of Zn in the GB massively increases during the segregation transition, markedly exceeding the previously observed large segregation transitions in iron alloys \cite{darvishikamachali2020segregation}.
For instance, at $T = 623$ K, about 60 at.\% Zn segregate to the GB if only 4 at.\% of Zn are added to BCC-Fe.

The critical alloy compositions, corresponding to the segregation transitions, can be obtained from segregation isotherms, i.e., the solid black line with squares on the left corner in Fig. \ref{fig_GBPhDia}(c), referred to as the \textit{transition line}. 
With this critical information, the GB phase diagram is complete:
Figure \ref{fig_GBPhDia}(c) illustrates that for alloy compositions and temperatures to the left side of the transition line, low segregation of Zn to the GB is expected. 
In contrast, shifting to the right side of the transition line induces massive Zn segregation, in which the GB composition dramatically increases, crossing its own miscibility gap.
This corresponds to the high-segregation branch in the segregation isotherms, Fig. \ref{fig_GBPhDia}(b). 
The results in Figs. \ref{fig_GBPhDia}(b) and (c) show that the magnitude of the segregation transition (the height of the jump) and the degree of segregation significantly increase with decreasing the temperature.
\revision{Such low temperatures and temperature decreases can occur during the cooling stages in galvanization and welding processes.}

The origin of the miscibility gap in the Fe-Zn bulk system has been attributed to a magnetic ordering effect, i.e., the presence of different magnetic states \cite{reumont2000thermodynamic,su2001thermodynamic}.
Using the density-based thermodynamic approach, here we show that this miscibility gap of the bulk is inherited by the GB and plays a central role in governing the segregation transition. 
To verify the existence of such a magnetic miscibility gap for GBs, we have performed DFT calculations of Zn segregation in a $\Sigma$5 GB, with and without the consideration of magnetic disorder.
The computational details are given in Appendix C in the SM. 
\revision{Figure \ref{fig_DFT_GB} reveals that the strong segregation for Zn to GBs discussed in previous DFT calculations \cite{bauer2015first,scheiber2020influence} is only present in the ferromagnetic case. 
Without magnetic ordering (paramagnetic state), the segregation of Zn only reduces the GB energy up to a coverage of 50 \%, i.e., the energy curve is convex.}
In the presence of a magnetic ordering (ferromagnetic state), the GB shows a concave energy curve, indicating that the Zn-segregated GB becomes thermodynamically unstable.
This demonstrates that the miscibility gap due to the magnetic ordering effect is inherited by the GB; thus, above a certain composition, the GB prefers to decompose into Zn-rich and Zn-poor segments.

\begin{figure}[t]
    \centering
   \includegraphics[width=1.0\linewidth]{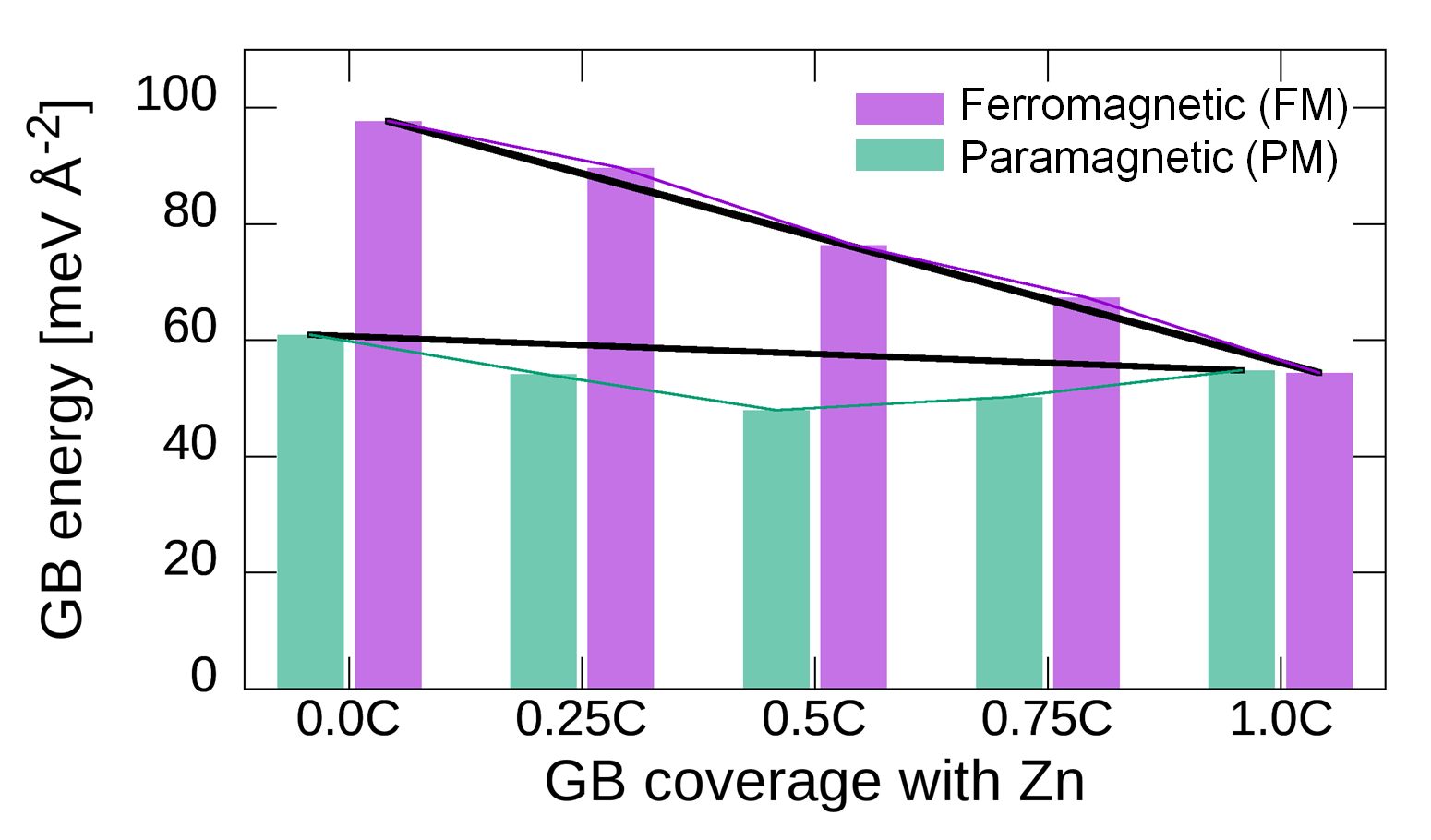}
    \caption{
    \textbf{GB energy from DFT calculations.}
    The energy of a $\Sigma$5 [100](013) GB is computed for various levels of Zn coverage (C: coverage), with and without the magnetic ordering effect.
    The black curves connect the GB energies for the end-members with no and full coverage.
    In the paramagnetic state, Zn favorably dissolves in the GB (convex) while in the ferromagnetic state, the GB shows a miscibility gap (concave).
    }
    \label{fig_DFT_GB}
\end{figure}
\begin{figure}[t]
    \centering
   \includegraphics[width=1.0\linewidth]{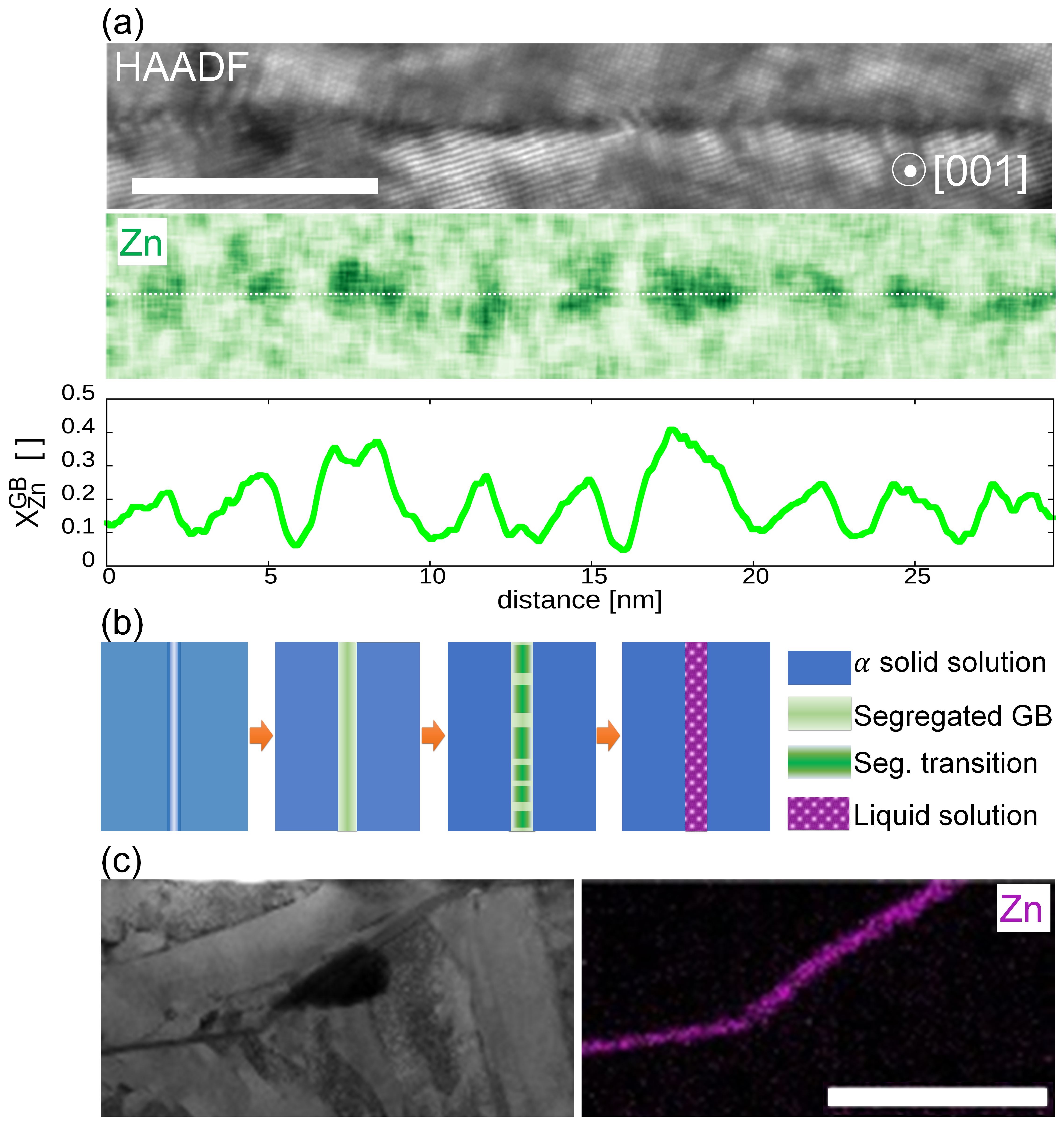}
    \caption{
    \textbf{GB segregation transition and related wetting.}
    (a) High-angle annular dark-field (HAADF) imaging in the scanning transmission electron microscope (STEM) showing a $\Sigma 5$ GB (scale bar 7 nm).
    Both grains are oriented along the [001] axis. 
    The corresponding elemental Zn map and line profile along the GB, obtained by STEM energy dispersive spectroscopy (STEM-EDS), reveal interfacial spinodal decomposition with localized Zn-rich patches, which is the signature of a segregation transition. 
    (b) Schematic sequence of segregation and segregation transition, assisting the formation of the liquid phase in the GB.
    (c) Bright-field scanning transmission micrograph (scale bar 1 $\mu$m) evidencing Zn-rich wetting along a GB after a spot welding process.
    Further experimental details are given in Appendix D in the SM.
    }
    \label{fig_Nucl}
\end{figure}
In fact, recent studies have revealed that such chemical decomposition at the GB is direct evidence of a segregation transition \cite{darvishikamachali2020segregation,wallis2023grain}.
To examine this for the Fe-Zn system, we performed high-resolution transmission electron microscopy of a $\Sigma 5$ GB in a Zn-coated BCC-Fe bicrystal.
Figure \ref{fig_Nucl}(a) shows a HAADF scanning transmission micrograph and the corresponding STEM-EDS elemental Zn map and line profile for the GB.
Experimental details are given in Appendix D in SM.
We found that with only $~$5 at.\% Zn in the bulk, the Zn strongly segregates to the GB and decomposes into rich and poor patches.
The existence of the Zn patches along the GB reveals an interfacial phase decomposition, as previously demonstrated \cite{darvishikamachali2020segregation,wallis2023grain}, and confirms the existence of the segregation transition in the Fe-Zn system. \revision{The concentration of Zn patches reaches 40 at.\% which is in the range of predicted values in Fig. \ref{fig_GBPhDia}(b). 
We note that an additional effect of gradient energy terms is neglected in the current calculations that can be responsible for the spread of Zn in the grain boundary region in Fig. \ref{fig_Nucl}(b).
Co-segregation with other elements is also evidenced to reduce the Zn segregation at the GB \cite{ahmadian2023interstitial}.}
The cross-validations from the DFT calculations and the experiment confirm the density-based modeling predictions of the segregation transition in the Fe-Zn system.

Given the complex temperature profiles experienced during the processing of Zn-coated steels \cite{hou2007finite}, the alloy compositions and experienced temperatures fall easily beyond the transition line in the GB phase diagram (Fig. \ref{fig_GBPhDia}(c)). 
\revision{Once the Zn supply into the alloy is available via diffusion into the grain interior, the GB segregation transition can be triggered.
This also can happen during the cooling stage, as the required amount of Zn in the grain to trigger the segregation transition is low and further decreases with decreasing the temperature.}
The experimental results shown in Fig. \ref{fig_Nucl}(a) reveal such segregation transition occurring during the cooling after a heat treatment process.
Interestingly, our calculations show that the Zn segregation becomes much stronger with decreasing the temperature.

The significant amount of Zn in the GB, originating from the segregation transition, can dramatically affect the course of GB wetting, as schematically shown in Fig. \ref{fig_Nucl}(b). 
Experimentally, we observed this in the welding process of a steel specimen, giving the thick Zn solution layer shown in Fig. \ref{fig_Nucl}(c). \revision{Although the segregation might not be the sole origin of wetting (e.g. stresses may play a significant role), the segregation transition is expected to dramatically boost it.} To understand this influence, we studied the barrier for the formation of the liquid phase from a Zn-enriched GB, $\Delta G=G^L-G^{GB}$.
Here the free energies of the liquid solution $G^L (T, X_{Zn})$ and the segregated GB, $G^{GB}=G(\rho = 0.92, T, X_{Zn})$ are obtained from CALPHAD database TCFE11 and through Eq. (\ref{eq_G_alloy}), respectively.
\begin{figure}[t]
    \centering
   \includegraphics[width=1.0\linewidth]{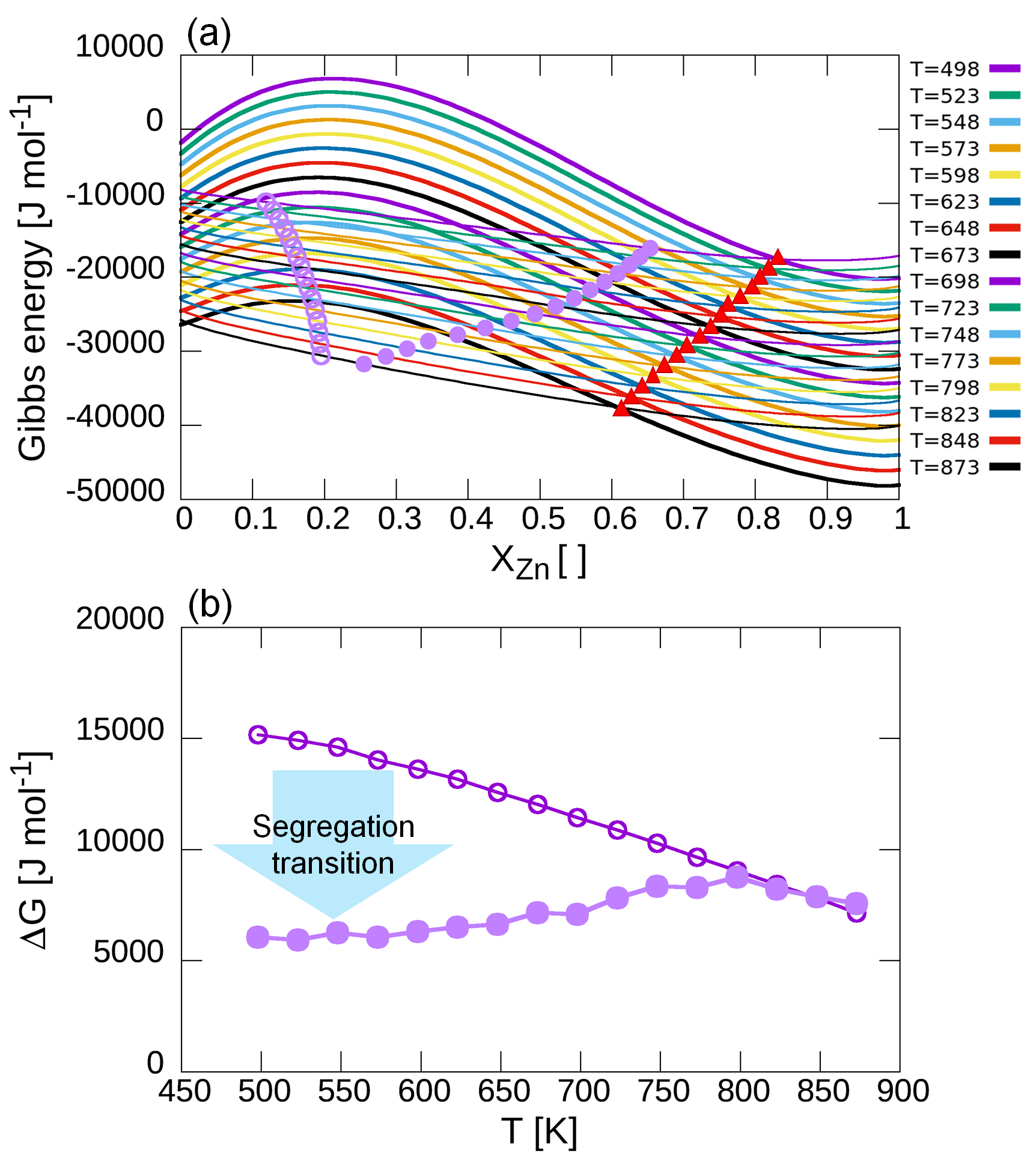}
    \caption{
    \textbf{Segregation-induced liquid formation at the GB.}
    (a) Free energies of GB (thin curves) and liquid Fe-Zn solution (thick curves) for different temperatures.
    The intersections of two free energy curves are also marked with triangles.
    For GB compositions beyond the intersection point, the liquid phase becomes stable, i.e., $\Delta G < 0$.
    GB compositions for right before and after the segregation transition are marked on the GB free energy curves with unfilled and filled circles, respectively. 
    (b) $\Delta G$ as a function of temperature are shown for the two sets of GB compositions marked in panel (a).
    The same markers are used as in panel (a).
    The arrow in panel (b) indicates that the massive segregation transition dramatically decreases the barrier for the formation of the liquid phase from segregated GBs.
    }
    \label{fig_Liquid}
\end{figure}

Figure \ref{fig_Liquid}(a) shows the free energies of the GB and the liquid solution for various temperatures.
The intersections of the two free energies corresponding to $\Delta G = 0$ are marked (triangles).
On the GB free energies, we also marked the GB compositions at the segregation transition, and the low and high values at the jump (circles).
We computed $\Delta G$ values for these two sets of GB compositions to specifically reveal the effect of segregation transition.
Figure \ref{fig_Liquid}(b) reveals that the barrier for the formation of the liquid phase significantly increases with a decrease in the temperature.
The results show that both compositional and energetic barriers for the formation of the liquid phase are greatly reduced by the Zn segregation transition.
\revision{Further applying the density-based model for intermetallic phases, we found that the segregation transition enables the formation of the Zn-rich $\Gamma$ phase ahead of the crack tip. This is thoroughly discussed in \cite{ikeda2023segregation}.}

\begin{figure}[b]
    \centering
   \includegraphics[width=1.0\linewidth]{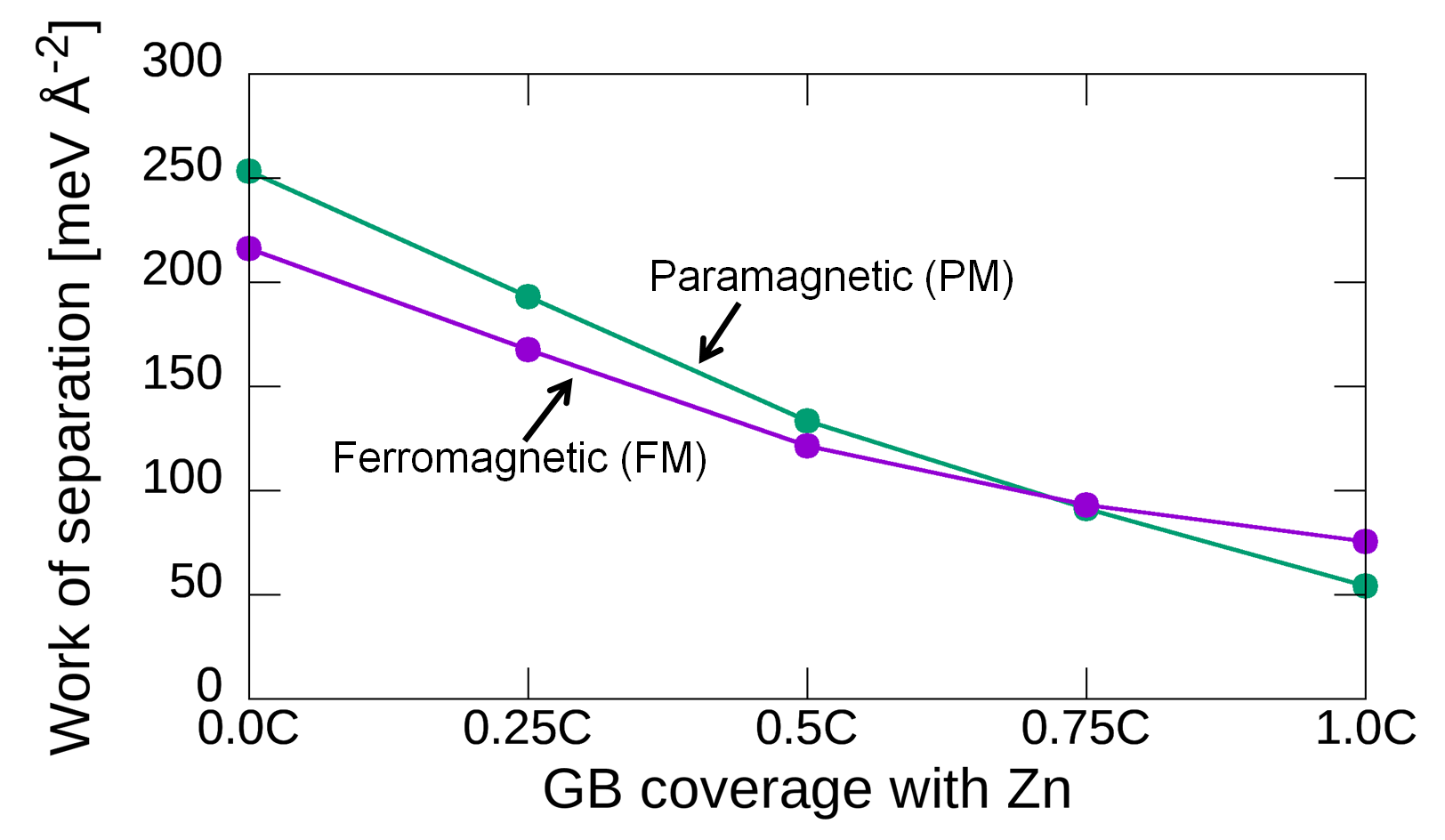}
    \caption{
    \textbf{GB's work of separation.}
    The work of separation of a $\Sigma$5 [100](013) GB for various levels of coverage with Zn, with and without magnetic ordering effect.
    }
    \label{fig_DFT_WofS}
\end{figure}

Prior to the liquid formation, the massive Zn segregation can also have great significance for promoting an easy crack initiation by weakening the GB.
We studied the work of separation of a $\Sigma$5 GB as a function of Zn coverage, using DFT calculations.
The results in Fig. \ref{fig_DFT_WofS} show that the segregation of Zn dramatically reduces the GB’s work of separation.
\revision{These confirm the results from previous works \cite{bauer2015first,scheiber2020influence,ahmadian2023interstitial}.}
We found that in the ferromagnetic state, the GB exhibits a slightly reduced work of separation and weaker dependence on the Zn content.
Nevertheless, the overall reduction in the work of separation is significant.
In fact, for 50\% Zn coverage, the work of separation for the GB reduces almost by half.
These results suggest that once the Zn segregation transition occurs, GB weakens dramatically.
Such a weakened GB ahead of a crack tip is highly prone to opening, thus promoting embrittlement in the material.

To summarize, we reveal and explain a giant segregation transition occurring during Zn segregation into Fe GBs using CALPHAD-integrated density-based modelling.
A central insight of the model is that it shows the GB can have a large magnetic miscibility gap, which is also confirmed by DFT calculations.
The segregation transition is evidenced by our HAADF-STEM experimental measurements, revealing an interfacial phase decomposition (formation of Zn-rich patches) along GB. 
Our results shed light on the origin and mechanisms of LME in Zn-coated steels and its strong dependence on temperature and alloy composition. 
In particular, we found that the amount of segregation dramatically increases with decreasing the temperature, while the required amount of Zn in the alloy to trigger the segregation transition decreases.
\revision{These emphasize the significance of the cooling stage during processing.}
We show that the segregation transition markedly assists with the formation of an interfacial liquid phase and weakens the GB by reducing the work of separation.
The strong temperature and composition dependence of the Zn segregation implies that LME in the Fe-Zn system should be preventable via alloy and processing design strategies that suppress the revealed segregation transition.
\revision{Follow-up density-based phase-field simulations and guided experiments are planned to explore the detailed influence of the alloying and processing parameters.}


%

\setlength{\parskip}{0pt}

\section*{Acknowledgments}
RDK acknowledges the financial support from the German research foundation (DFG) within the project \emph{DA 1655/3-1} and the Heisenberg program’s project \emph{DA 1655/2-1}.
US and TH acknowledge the DFG support within the transfer project T07 of the \emph{SFB761} ``Stahl - ab initio''. 
CHL acknowledges the DFG support within the project \emph{LI 2133/7-1.}
YI gratefully acknowledges the Takenaka Overseas Scholarship Foundation for their kind support of his graduate work. 
This research was carried out in part at the electron microscopy center at BAM.

\section*{Author contribution}
RDK conceived the idea and collaborations, conducted the thermodynamic calculations, supervised grain boundary density calculations and wrote the manuscript.
TW computed the grain boundary density from atomistic simulations.
US conducted the ab initio calculations.
AA conducted the experiments on the bicrystal specimen.
YI conducted the experiments on the welding specimen.
TH supervised the ab initio calculations and revised the manuscript.
CHL supervised the experiments on the bicrystal specimen and revised the manuscript.
RM supervised the experiments on the welding specimen and revised the manuscript.
All authors reviewed and edited the manuscript.
RDK, CHL, TH and RM secured funding.

\section*{Supplementary material}
Methods and additional discussions are presented in the Supplementary Material (SM) attached.

\section*{Data availability}
Further data and codes of this study are available from the corresponding author.

\section*{Competing financial interests}
The authors declare no competing financial interests.



\section*{Supplementary material}

Supplementary Material for
# Giant segregation transition as origin of liquid metal embrittlement in the Fe-Zn system

Reza Darvishi Kamachali[*,a], Theophilus Wallis[a], Yuki Ikeda[a], Ujjal Saikia[b], Ali Ahmadian[b], Christian H. Liebscher[b], Tilmann Hickel[a, b], Robert Maaß[a, c]

[a] *Federal Institute for Materials Research and Testing (BAM), Unter den Eichen 87, 12205 Berlin, Germany*
[b] *Max-Planck-Institut für Eisenforschung GmbH, Max-Planck-Straße 1, 40237 Düsseldorf, Germany*
[c] *Department of Materials Science and Engineering, University of Illinois at Urbana-Champaign, Urbana, Illinois 61801, USA*

In this supplementary document, additional information on CALPHAD-integrated density-based modelling as well as related ab initio calculations and experimental measurements are presented. All figures are vividly described in their corresponding caption. Specific points of discussion can be found in the main paper. In the following:

- **Figure S1: The miscibility gap in Fe-Zn phase diagram**
- **Appendix A: Calculation of grain boundary density**
- **Appendix B: CALPHAD-integrated density-based modelling**
- **Appendix C: Ab initio calculations**
- **Appendix D: Experimental measurements**

---

[*]Corresponding author: reza.kamachali@bam.de (Reza Darvishi Kamachali)

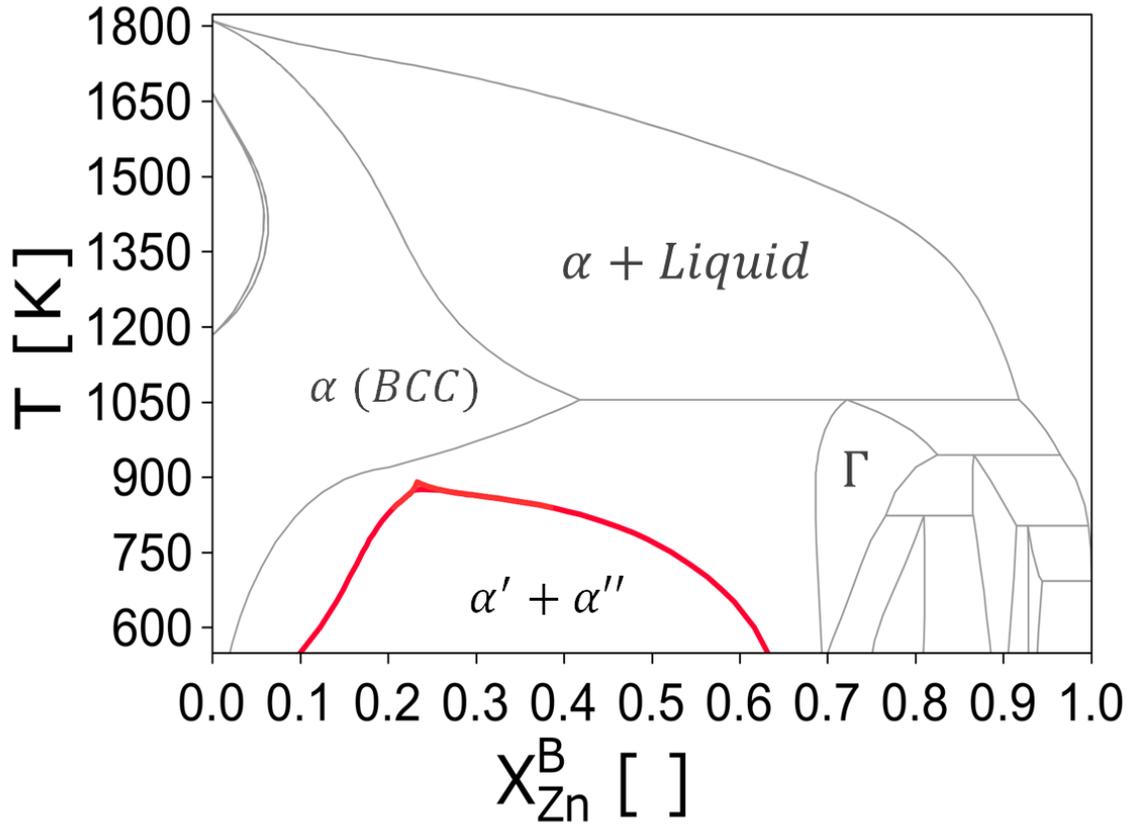

**Figure S1: The miscibility gap in Fe-Zn phase diagram.** A miscibility gap (red curve) is predicted for the BCC Fe-Zn which is shown to originate from the magnetic ordering effect [1, 2]. With this miscibility gap, the BCC Fe-Zn solid solution ($\alpha$) decomposes into Zn-lean ($\alpha'$) and Zn-rich ($\alpha''$) phases, spreading over a wide range of temperatures and compositions. The same miscibility gap is shown in Fig. 1(a) to compare with the GB miscibility gap computed from CALPHAD-integrated density-based method. The superscript B in $X_{Zn}^B$ indicates 'bulk'. Thermo-Calc TCFE11 database is used for computing the phase diagram and the embedded miscibility gap.

# Appendix A: Calculation of grain boundary density

For our calculations, we need to have a realistic value for the average GB density. For this purpose, the equilibrium GB structures obtained from the atomistic simulations [3] were processed to compute the continuous atomic density profile on the mesoscale. A coarse-graining approach has been implemented in which the delta functions, representing the atoms, in

$$\rho(q) = \sum_I \delta(q - A_I) \tag{1}$$

are replaced with a Gaussian distribution function

$$\delta(q) = \frac{e^{\frac{-(q-A_I)^2}{2\beta^2}}}{\beta\sqrt{2\pi}} \tag{2}$$

with $q$ the spatial coordinate in the simulation domain and $A_I$ the position of atom $I$. The coarse-graining length $\beta$ is related to the interatomic interaction distance, here chosen to be $\beta = 1.4 r_{Fe}$, with $r_{Fe} = 1.26$ Å, the atomic radius of Fe. By normalizing the density profiles obtained from the coarse-graining approach with the bulk atomic density, we then obtain the dimensionless atomic density across the GB.

Several $\Sigma$ GBs in BCC-Fe with relatively high GB energies were studied. Figure S2 shows the results, including (i) atomic structure, (ii) cross-section view of the atomic density field and (iii) the corresponding atomic density profiles for the GBs, respectively. The error bars in the density profiles show the in-plane density fluctuations. The results show that the average GB density has a finite range of $\sim [0.92 - 0.94]$, with fluctuations over $\sim [0.88 - 0.98]$. Based on these results, we chose the average GB density value $\rho = 0.92$ for the CALPHAD-integrated density-based investigations. Table 1 lists relevant data for the GBs shown in Fig. S2, including the GB energy $\gamma^{GB}$, the coincidence site lattice value $\Sigma$, the density values and the GB excess free volume $\Delta V$.

Table S1: Relevant GB data for the GBs shown in Fig. S2.

| GB index | $\gamma^{GB}$ [J/m²] | $\Sigma$ | ave. GB density | Fluc. in GB density | $\Delta V$ [Å] |
|---|---|---|---|---|---|
| 1 | 1.09 | 5 | 0.925 | [0.909 - 0.941] | 0.344 |
| 2 | 0.99 | 5 | 0.924 | [0.920 - 0.928] | 0.314 |
| 3 | 1.07 | 17 | 0.933 | [0.881 - 0.984] | 0.278 |
| 4 | 1.06 | 3 | 0.938 | [0.931 - 0.945] | 0.265 |
| 5 | 1.15 | 9 | 0.934 | [0.926 - 0.943] | 0.329 |
| 6 | 1.14 | 5 | 0.923 | [0.890 - 0.948] | 0.346 |
| 7 | 1.10 | 17 | 0.938 | [0.895 - 0.982] | 0.295 |
| 8 | 1.17 | 29 | 0.924 | [0.892 - 0.952] | 0.324 |
| 9 | 1.27 | 3 | 0.935 | [0.923 - 0.947] | 0.344 |
| 10 | 1.11 | 5 | 0.928 | [0.887 - 0.950] | 0.343 |
| 11 | 1.16 | 5 | 0.927 | [0.898 - 0.958] | 0.338 |
| 12 | 1.28 | 49 | 0.926 | [0.901 - 0.947] | 0.344 |

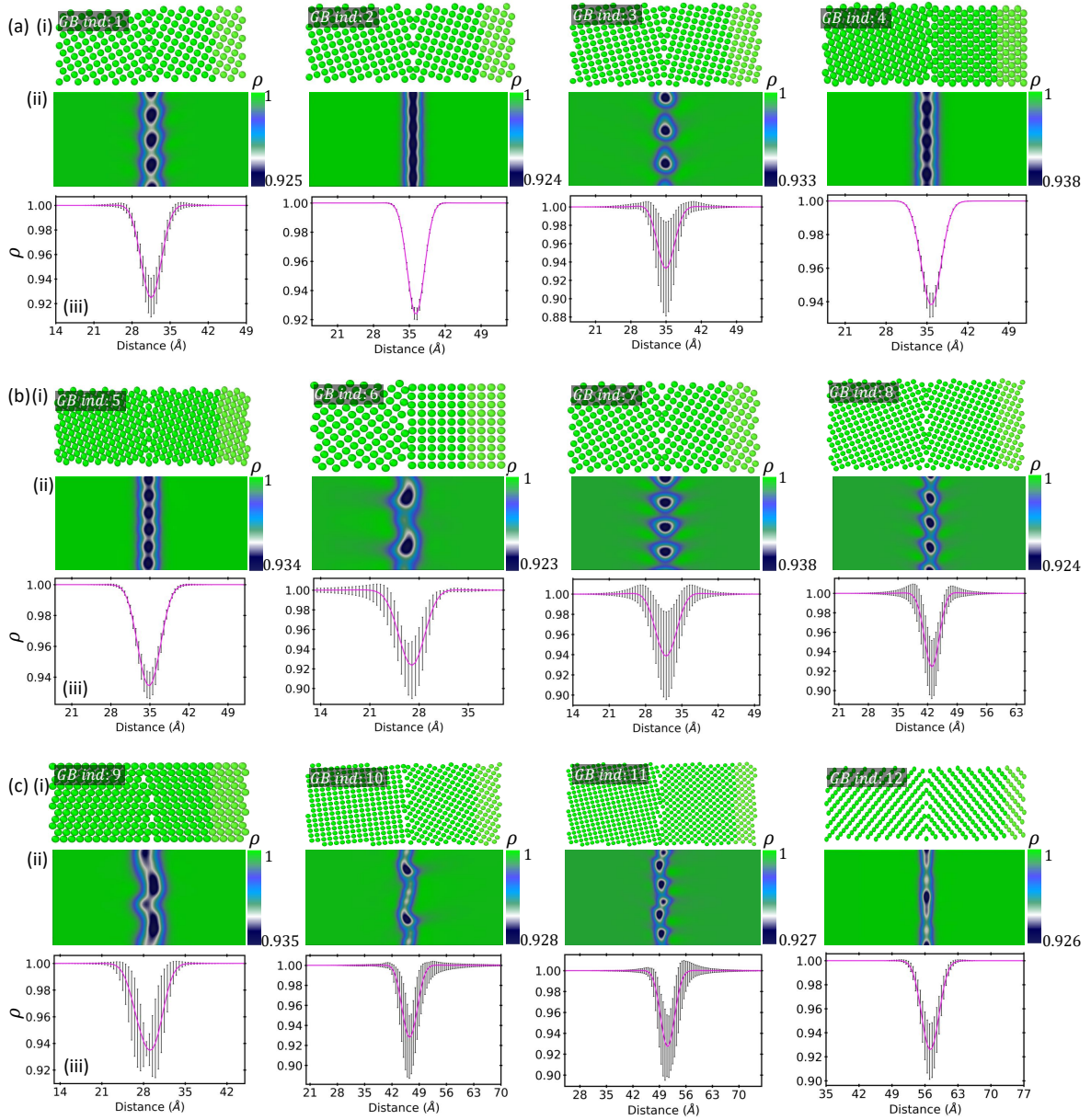

**Figure S2: The atomic structures and the coarse-graining results for several GBs.** For each GB, (i): the equilibrium atomistic GB structure in cross-section view, (ii): the coarse-grained continuous density field for the corresponding GB, scaled with the bulk density and (iii): the corresponding atomic density profiles are shown. The error bars represent the in-plane fluctuation in the atomic densities.

# Appendix B: CALPHAD-integrated density-based modelling

The density-based thermodynamic model has been proposed to accelerate assessment and investigation of GB segregation and phase behavior [4]. In this model, a mesoscale atomic density $\rho$ that describes the GB region with respect to the bulk is used to approximate GB Gibbs free energy. This is integrated with CALPHAD framework such that the bulk thermodynamic data available from CALPHAD databases are then directly used as input. Recently, CALPHAD-integrated density-based method has been successfully applied to predict GB properties and behavior in a broad range of materials [5–11]. For the binary Fe-Zn system, the density-based Gibbs free energy can be written as

$$G(T, \rho, X_{Zn}) = X_{Fe} G_{Fe}(T, \rho) + X_{Zn} G_{Zn}(T, \rho)$$
$$+ \rho^2 \Delta H_{mix}^B(T, X_{Zn}) - T \Delta S_{mix}^B(T, X_{Zn})$$
$$+ \frac{\kappa_{X_{Zn}}}{2}(\nabla X_{Zn})^2$$

with

$$G_i(T, \rho) = \rho^2 E_i^B + \rho \left( G_i^B(T) - E_i^B \right) + \frac{\kappa_i^\rho}{2} (\nabla \rho)^2. \tag{3}$$

Here $X_{Zn}$ is the molar fraction, $T$ the temperature, $G_i^B$ and $E_i^B$ the Gibbs free energy and potential energy of the pure bulk substance $i$ and, $\Delta H_{mix}^B$ and $\Delta S_{mix}^B$ are the mixing enthalpy and entropy of the bulk solid solution. The superscript $B$ indicates the bulk thermodynamic data to be retrieved from CALPHAD database. $\kappa_{Zn}$ and $\kappa_i^\rho$ are the concentration and density gradient energy coefficients, respectively. The gradient energy terms account for the chemical and structural heterogeneities in the GB region. Assuming a constant average GB density, we simplified our modelling considerations by neglecting the gradient energy terms. Instead, in this study, we focus on the temperature and composition dependence of the GB segregation and phase behavior and their implications for the liquid metal embrittlement (LME) phenomenon.

We consider GB density $\rho = 0.92$, deduced from analyzing the results of atomistic simulations presented in Appendix A. First, in order to obtain the GB miscibility gap, we computed the isothermal Maxwell construction. The Maxwell construction is equivalent to the common tangent construction and can be computed on the chemical potential curves as described in Ref. [4]. In order to perform the calculations, a C++ code was developed. The Thermo-Calc TCFE11 database has been used to retrieve the bulk thermodynamic data, over the temperature range of 498–873 K. The potential energies of Fe and Zn are calculated as described in Ref. [6], to be $-77.2$ and $-24.3$ kJ mol$^{-1}$, respectively. The resulting GB miscibility gap is shown in Fig. 1(a) in the main paper. Further, in order to obtain the segregation isotherms, we applied the equal chemical potential condition between the bulk and GB. Theoretically, this should give the Fowler–Guggenheim relation for segregation isotherm, as discussed in Ref. [4]. However, since the analytical description of the mixing terms $\Delta H_{mix}^B$ and $\Delta S_{mix}^B$ were not available (we only use the values retrieved from the database), we applied a numerical method, such that, for any given bulk composition $X_{Zn}^B$, the corresponding chemical potential and, subsequently, the corresponding GB composition $X_{Zn}^{GB}$ for the same chemical potential were computed. For these calculations, the composition and chemical potential were resolved with accuracies better than $10^{-4}$ and 100 J mol$^{-1}$, respectively. The resulting segregation isotherms are shown in Fig. 1(b) in the main paper. Further computational details about the density-based model can be found in Refs. [4–6].

To compute the driving force for the formation of the liquid phase, we retrieved the free energy of the liquid Fe-Zn solution $(G^L)$ from the Thermo-Calc TCFE11 database and the free energy of the Zn-segregated GB is obtained using Eq. (3). The corresponding free energy curves for various temperatures are shown in Fig. 4 (a). At a given temperature $T$ and GB composition $X_{Zn}^{GB}$, the driving force energy is computed by

$$\Delta G(T, X_{Zn}^{GB}) = G^L(T, X_{Zn}^{GB}) - G(T, \rho = 0.92, X_{Zn}^{GB}) \tag{4}$$

In our calculations presented in Fig. 4(b), we considered two sets of GB compositions, corresponding to the segregation transition, one for right before the transition and one for right after. The values are listed in Table S2. Note that these values also mark the GB miscibility gap and correspond to the alloys on the transition line shown on the GB phase diagram, Fig. 1(c) in the main paper, as well. The computed $\Delta G$ values are shown in Fig. 4(c), indicating the effect of GB segregation transition on the formation of the liquid phase from the Zn-enriched GB.

Table S2: GB compositions at the segregation transition jump.

| T [K] | $^-X_{Zn}^{GB}$ | $^+X_{Zn}^{GB}$ |
|---|---|---|
| 498 | 0.11865 | 0.65365 |
| 523 | 0.12855 | 0.63965 |
| 548 | 0.13685 | 0.62465 |
| 573 | 0.14405 | 0.60835 |
| 598 | 0.15035 | 0.59035 |
| 623 | 0.15615 | 0.57025 |
| 648 | 0.16175 | 0.54765 |
| 673 | 0.16705 | 0.52205 |
| 698 | 0.17235 | 0.49295 |
| 723 | 0.17755 | 0.46025 |
| 748 | 0.18245 | 0.42405 |
| 773 | 0.18685 | 0.38525 |
| 798 | 0.19005 | 0.34495 |
| 823 | 0.19205 | 0.31535 |
| 848 | 0.19365 | 0.28565 |
| 873 | 0.19505 | 0.25485 |

## Appendix C: Ab initio calculations

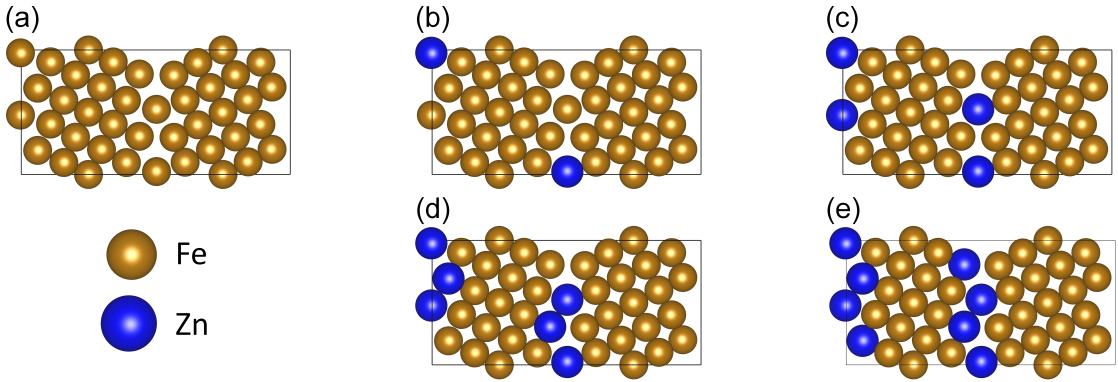

**Figure S3: DFT simulated GBs.** Σ5 [100](013) GB structure used in the DFT calculations is shown for pure iron and different levels of Zn coverage.

The first principles calculations were performed with density functional theory (DFT) [12] as implemented in the Vienna Ab-initio Simulation Package (VASP) [13]. We used projector-augmented wave (PAW) potentials [14] with the generalized gradient approximation of Perdew-Burke-Ernzerhof (PBE) [15]. A kinetic energy cutoff of 400 eV was for plane waves with a k-point sampling of 18×18×18 for Fe and 28×28×28 for Zn was used. With these parameters we obtained bulk lattice constant values of 2.832 Å and 2.831 Å for bcc FM and bcc PM Fe respectively. Ionic relaxations

6

were performed until the magnitude of the force components on all atoms were less than 0.01 eV/Å. We used super-cell approach to construct the GBs and surfaces with periodically repeated orthorhombic cells.

We calculate the interface energy of the pristine GBs and surfaces as

$$\gamma_{GB} = \frac{E_{Fe:GB} - N_{Fe}\mu_{Fe}}{2A} \quad (5)$$

$$\gamma_{surf} = \frac{E_{Fe:surf} - N_{Fe}\mu_{Fe}}{2A} \quad (6)$$

Here, $E_{Fe:GB}$ is the DFT total energy obtained from an explicit GB simulation and $N_{Fe}$ the number of iron atoms exchanged between the GB and the iron reservoir, characterized by its chemical potential $\mu_{Fe}$. The factor $1/2$ accounts for the presence of two GBs within our simulation cell, a consequence of the periodic boundary conditions we used and $A$ is area of the GB interface plane. Similarly, $E_{Fe:surf}$ is the total energy of the surface slab. The interface energy of Zn covered GBs and surfaces are calculated as

$$\gamma_{GB} = \frac{E_{Zn@Fe:GB} - N_{Fe}\mu_{Fe} - N_{Zn}\mu_{Zn}}{2A} \quad (7)$$

$$\gamma_{surf} = \frac{E_{Zn@Fe:surf} - N_{Fe}\mu_{Fe} - N_{Zn}\mu_{Zn}}{2A} \quad (8)$$

Here, $E_{Zn@Fe:GB}$ is the total energy of the Zn covered GB and $N_{Zn}$ the number of iron atoms exchanged between the GB and the Zn reservoir, characterized by its chemical potential $\mu_{Zn}$. The term, $E_{Zn@Fe:surf}$ is the total energy of the Zn covered surface slab.

The Zn induced embrittling behavior was investigated by calculating the work of separation as

$$W_{sep} = 2\gamma_{surf}(\mu_{Zn}) - \gamma_{GB}(\mu_{Zn}) \quad (9)$$

In this study, we adopted spin-space averaging (SSA) technique to describe paramagnetic GBs [16]. For atomic relaxations of the paramagnetic structures, we used SSA relaxation approach. The core approximation of this approach is that magnetic configurations changes so fast that an atom cannot respond to a single magnetic configuration. Hence, instead of instantaneous force of each magnetic configuration, the atoms move according to the averaged force over different magnetic configurations.

## Appendix D: Experimental measurements

For the experiments, two materials were considered. First, for studying Zn segregation and the corresponding GB phase decomposition phenomena, an Fe-4at.%Al bicrystal was used, with a global $\Sigma 5\,[0\,0\,1]\,(3\,1\,0)$ GB separating the two grains. The bicrystal was cut into a $\sim 15 \times 11 \times 1\,mm$ rectangle, mechanically grinded, chemically polished and then immersed into a liquid Zn bath of 99.999% purity for 300 s at 740 K. The Zn-coated sample was grinded again and thereafter heat treated at 1073 K for 80 hours and then air-cooled to room temperature. In order to study the GB, TEM plane-view specimen was prepared from the area beneath the layer of Fe-Zn intermetallic phases. HAADF-STEM imaging as well as STEM-EDS were performed on a $C_s$ probe-corrected FEI Titan Themis $60-300\,kV$ equipped with the SuperX EDS detector system. The investigations were done with a semiconvergence angle of $17\,mrad$, a camera length of $100\,mm$ and a probe current of $80\,pA$. While the images were acquired at $300\,kV$, the high-resolution STEM-EDS mapping was recorded at $120\,kV$. The corresponding results are shown in Fig. 3(a) in the main paper.

Second, an industrially produced 1.4 mm thick third-generation galvanized (GI) advanced high-strength steel (AHSS) sheet, containing a maximum of 0.26 % C, 2.0 % Si, 2.3 % Mn, 0.04 % P, 0.01 % S, 1.0 % Al, 0.2 % Cu, 0.005 wt. % B, 0.15 %(Ti + Nb), and 0.6 % (Cr + Mo) (all in wt.%)

and with a tensile strength exceeding 1 GPa, was used to study thick Zn-rich layers forming during welding. The AHSS was annealed in $N_2$ -5 %$H_2$ atmosphere containing a partial pressure, $pO_2$, of $2.5210^{-21}$ atm, followed by galvanization in a Zinc - Aluminum bath, which produced a GI-coating on top of the steel. The GI-AHSS was resistance spot welded (RSW) in a three-layer stack assembly, where the top sheet was the GI-AHSS, followed by two 1.5 mm thick sheets of GI extra deep drawing steel (EDDS). RSW was carried out using a 75/85kVA Taylor-Winfield RSW instrument with a 6 mm diameter copper electrode. The welding condition was chosen according to the SEP1220-2:2011-08 guidelines. During RSW, a constant electrode force of 4.5 kN and current of 9.2 kA were applied. The applied welding current is just below expulsion, corresponding to the highest current for achieving a maximum nugget size. For the here investigated samples, the duration of the welding squeeze, welding time, and hold time were 1167 ms, 250 ms, and 300 ms, respectively. Following welding, cross-sectional samples of the center of the weld were prepared. These were mechanically polished according to standard procedures, ending with a chemo-mechanical colloidal silica (50 nm particle size) polish. TEM JEM-2200FS with EDS detector was used for detailed structural investigations. TEM specimens were prepared using focused ion beam (FIB) Quanta 3D FEG and extracted from the steel in an in-plane orientation. The results are shown in Fig. 3(c) in the main manuscript.



\end{document}